\documentclass[12pt,preprint]{aastex}

\begin{document}
\shorttitle{SN 2004dj's Light Curve}

\shortauthors{Zhang, T. et al.}

\title{Optical Photometry of Type II-P Supernova 2004dj in NGC 2403}

\author{Tianmeng Zhang\altaffilmark{1}, Xiaofeng Wang\altaffilmark{ 1,2},
Weidong Li\altaffilmark{3}, \\ Xu Zhou\altaffilmark{1}, J un
Ma\altaffilmark{1}, Zhaoji Jiang\altaffilmark{1}, Jiansheng
Chen\altaffilmark{1}}

\altaffiltext{1}{National Astronomical Observatories of China,
Chinese Academy of Sciences, Beijing 100012, P.R. China;
ztm@vega.bac.pku.edu.cn, wxf@vega.bac.pku.edu.cn}
\altaffiltext{2}{Physics Department and Tsinghua Center for
Astrophysics, Tsinghua University, 100084, Beijing, China;
wang\_xf@mail.tsinghua.edu.cn} \altaffiltext{3}{Department of
Astronomy, University of California, Berkeley, CA94720-34 11;
weidong@astron.berkeley.edu}

\begin{abstract}

We present photometric data of the type II-P supernova (SN)
2004dj in NGC 2403. The multicolor light curves cover the SN from
$\sim$ 60 to 200 days after explosion, and are measured with a set
of intermediate-band filters that have the advantage of tracing
the strength variations of some spectral features. The light
curves show a flat evolution in the middle of the plateau phase,
then decline exponentially at the late times, with a rate of
0.10$\pm$0.03 mag (10 days)$^{-1}$ in most of the filters. In the
nebular phase, the spectral energy distribution (SED) of SN 2004dj
shows a steady increase in the flux near 6600~\AA\, and 8500~\AA,
which may correspond to the emission lines of  H$\alpha$ and Ca~II
near-IR triplet, respectively. The photometric behavior suggests
that SN 2004dj is a normal SN II-P. Compared with the light curves
of another typical SN II-P 1999em, we estimate the explosion date
to be June 10$\pm$21 UT, 2004 (JD 2453167$\pm$21) for SN 2004dj.
We also estimate the ejected nickel mass during the explosion to
be $M(^{56}\rm{Ni})$ = 0.023 $\pm$ 0.005 $M_{\odot}$ from two
different methods, which is typical for a SN II-P. We derive the
explosion energy $E \approx 0.75^{+0.56}_{-0.38}\times10^{51}$
erg, the ejecta mass $M \approx 10.0^{+7.4}_{-5.2}$ $M_{\odot}$, and
the initial radius $R \approx 282^{+253}_{-122}$ $R_{\odot}$ for the
presupernova star of SN 2004dj, which are consistent with other
typical SNe II-P.

\end{abstract}


\section{Introduction}

Supernova (SN) 2004dj was discovered by K. Itagaki \citep{nak04}
on 2004 July 31.76 (UT dates are used throughout this paper) in
the nearby SBcd galaxy NGC 2403 at a distance of 3.3$\pm$0.1 Mpc
\citep{kar04}. With a peak magnitude of 11.2 mag in the $V$ band,
SN 2004dj was the brightest supernova in the past decade.
Spectroscopic observations of SN 2004dj show that it is a typical
Type II Plateau SN (SN II-P), with prominent P-Cygni profiles in
hydrogen Balmer lines \citep{pat04}. The main photometric
characteristic of a SN II-P is that its light curve, unlike other
SNe, does not decay rapidly after maximum, but shows a plateau
phase for 60-100 days before decaying exponentially. The plateau
phase originates from a balance between the receding photosphere
in the expanding ejecta when the supernova is powered by the
recombination of hydrogen previously ionized by the supernova
shock. SNe II-P have long been thought to be produced by
core-collapse of massive ($>$8 $M_{\odot}$) red supergiants that do
not experience significant mass loss and retain  most of their
hydrogen-rich envelopes.

The optical position of SN 2004dj is measured to be R. A. =
07$^{\rm{h}}$37$^{\rm{m}}$17$^{\rm{s} }$.04, Dec =
+65$^{\circ}$35$^{\prime}$57$^{\prime\prime}$.84 (J2000.0;
\citet{nak04}). This is in good agreement with the {\it Chandra}
X-ray position \citep{poo04} and the  MERLIN radio position
\citep{bes05}. Using the astrometry of SN 2004dj, and also from
geometrical registration between images of NGC 2403 before and
after SN 2004dj occurred, both \citet{mai04} and \citet{wxf05}
have convincingly shown that SN 2004dj occurred at a position
coincident with Sandage Star 96 (hereafter S96) in the list of
luminous stars and clusters in NGC 2403 published by
\citet{san84}. \citet{mai04} suggest that S96 is a young compact
cluster with an age of 13.6 Myr and a total stellar mass of
$\sim$ 24,000 $M_\odot$, and estimate that the progenitor of SN
2004dj had a main-sequence mass of about 15 $M_\odot$.
\citet{wxf05}, on the other hand, suggest that S96 is an older
($\sim$ 20.0 Myr) and more massive ($\sim$ 96,000 $M_\odot$)
cluster, and that the progenitor of SN 2004dj may have a mass of
$\sim$ 12 $M_\odot$.

Studies of SN 2004dj itself may shed light  on the nature of its
progenitor, thus providing additional constraints on the
properties of S96. In this paper, we present the results from our
campaign of photometric followup of SN 2004dj from Aug 2004 to Jan
2005. The observations and data reduction are described in $\S$ 2,
and the multicolor light curves and spectral energy distribution
(SED) are presented in $\S$ 3. We estimate the explosion date, the
reddening, the synthesized $^{56}$Ni mass, and some explosion
parameters for SN 2004dj in $\S$ 4. The conclusions are summarized
in $\S$ 5.

\section{Observations and Data Reduction}

The observations of SN 2004dj were conducted with the 60/90 cm
Schmidt telescope located at the Xinglong station of the National
Astronomical Observatory of China (NAOC). A Ford Aerospace
2048$\times$2048 CCD camera with a 15 micron pixel size is mounted
at the Schmidt focus of the telescope. The field of view of the
CCD is $58'\times58'$ with a pixel scale of  1.7$\arcsec$.

This telescope has a photometric system with 15 intermediate-band
(FWHM $\approx$ 200-400~\AA) filters covering a wavelength range
from 3000~\AA\ to 10000~\AA. Figure 1 shows the filters
transmissions. These filters, also dubbed as the  BATC system,
were used to conduct a survey project among astronomers in
Beijing, Arizona, Taipei, and Connecticut \citep{fan96}, and are
designed to avoid contamination from the strongest and most
variable night sky emission lines. The BATC magnitudes use the
monochromatic AB magnitudes as defined by \citet{ok83}. The
standard stars HD 19445, HD 84937, BD+262606, and BD+174708
\citep{ok83} are observed for flux calibration in the BATC
observations. A full description of the BATC photometric system
can be found elsewhere \citep{fan96, yan99, zhou03}.

We began to monitor SN 2004dj shortly after its discovery. The
observations were obtained with 12 out of the 15 intermediate-band
filters, and covered a period from 11 Aug 2004 to 10 Jan 2005.
Typical exposure times are 300 s for each of the filters (longer
for filters with the shortest wavelengths).

\subsection{Photometry}

As demonstrated by \citet{mai04} and \citet{wxf05}, SN 2004dj
occurred in a compact star cluster (S96) with  $V \sim 18.0$ mag.
The background of SN 2004dj is also contaminated by the spiral
arms and nearby H~II regions in NGC 2403. To perform proper
photometry for SN 2004dj, subtraction of a template image of NGC
2403 without the SN is necessary. Usually observers have to wait
for a long time until a SN fades enough to take the template image
of its host galaxy, but fortunately for us, the host galaxy of SN
2004dj has been imaged as part of the Multicolor Sky Survey of
BATC since 1995. The enormous archival images of NGC 2403
\citep{wxf05} provide us with the necessary template images for
all the filters.

The goal of the image subtraction is to remove the contamination
of the SN 2004dj flux by the compact cluster itself, and by the
underlying emission from the spiral arms and nearby H~II regions.
To perform the subtraction, the observation with SN 2004dj is
first geometrically registered to the corresponding template
image. The intensity of the two images are then matched by
comparing the total fluxes of the stars in the images, and the
point-spread-function (PSF) of the two images are convolved to
the same level. Finally, the template is subtracted from the SN
image, leaving a clean background for the SN. Figure 2
demonstrates this process.

The final step is to perform standard aperture photometry on the
subtracted images. We used Pipeline II (a program developed to
measure the magnitudes of point sources in BATC images) that is
based on Stetson's DAOPHOT package \citep{ste87}.

\subsection{Calibration}

A total of 45 photometric nights were used to calibrate 20 local
standard stars in the field of SN 2004dj. For each photometric
night, the afore-mentioned four standard stars were observed in a
range of airmasses, and we derive iteratively the extinction
curves and the slight variation of the extinction coefficients
with time ($K+\Delta (UT)$) \citep{zhou01}. The instrumental
magnitudes ($m_{inst}$) are transformed to the BATC AB magnitude
($m_{\rm{batc}}$) \citep{zhou01} by

\begin{equation}
m_{\rm{batc}}=m_{\rm{inst}}+[K+\Delta (UT)]\chi + C.
\end{equation}
where $\chi$ is the airmass, and $C$ is the zero point of
magnitude.

Table 1 lists the final calibrated BATC magnitudes and their
uncertainties of the 20 local standard stars. Figure 3 shows a
$30'\times30'$ image of the SN 2004dj field, together with the
local standard stars within this frame.

These local standard stars are then used to transform the
instrumental magnitudes of SN 2004dj to the BATC system, and the
results are listed in Table 2. The estimated error shown in the
parenthesis is a quadrature sum of the uncertainties in the bias
and flat-field correction, the aperture photometry, and the
calibrations. The main source of the error comes from the
photometry, which is mainly caused by photon noises and the
uncertainties in the image subtraction.

The last three columns of Table 2 list the broad-band
Johnson-Cousins $VRI$ magnitudes transformed from the BATC
magnitudes. This transformation is done using the formula
established by \citet{zhou03} which is given here as,
\begin{eqnarray}
U&=&b+0.6801(a-b)-0.8982\pm0.143,\nonumber\\
B&=&d+0.2201(c-d)+0.1278\pm0.076,\nonumber\\
V&=&g+0.3292(f-h)+0.0476\pm0.027,\\
R&=&i+0.1036\pm0.055,\nonumber\\
I&=&o+0.7190(n-p)-0.2994\pm0.064.\nonumber
\end{eqnarray}
We need to point out, however, these transformation equations are
derived from observations of Landolt standard stars, and may have
relatively large uncertainties when applied to objects with broad
lines, such as SNe.

\section{Multicolor Light Curves}

\subsection{Light Curves of the first 200 days}

The light curves of SN 2004dj in twelve intermediate-band filters
(from $d$ to $p$) are shown in Figure 4. Our photometry started
at JD 2453230 and ended at JD 2453377, which corresponds to
$\sim$ 60 to 200 days after explosion if we adopt 2004 June 10 (JD
2453167) as the day of explosion for SN 2004dj (see the analysis
in \S 4.1). Our observations show that SN 2004dj has a prominent
plateau phase in all the passbands, suggesting that SN 2004dj is a
SN II-P.

The light curves in all bands were characterized by a phase of 45
days of slowly decreasing luminosity (until JD 2453275). During
this phase, the magnitude of SN 2004dj decreased by slightly
different amount for different bands: 0.20 mag in $n$, 0.65 mag in
$d$ and $g$, and $\lesssim$ 0.5 mag in the other bands.

After the plateau phase, the light curves show a rapid drop of
$\sim$ 2 mag in about 30 days, followed by a linear decay in
magnitude that signals the onset of the nebular phase. The slow
luminosity decline at the nebular phase, as listed in Table 3, is
evident for SN 2004dj. The decline rates in most bands are
$0.10\pm0.03$ mag (10 days)$^{-1}$ during the epochs from $\sim$
130 days to about 200 days after explosion. These late-time decay
rates are in good agreement with the radioactive decay of
$^{56}$Co.

The evolution of the nebular-phase light curves in the $i$ and
$n$ bands do not follow the other bands. Instead, the $i$-band
flux remains almost constant, while the $n$-band flux increases
slowly during the nebular phase. The peculiar behavior of the SN
in these two bands may be due to the effect of strong emission
lines, as discussed in the next section.

For comparison the light curves of SN 2004dj in the $VRI$ band,
transformed from our intermediate-band light curves using Eq.(2),
are plotted along with the typical type II-P SN 1999em
\citep{leo02, ham01} in Figure 5. Also plotted in the figure is
the $V$-band light curve of SN 2004dj as reported by
\citet{chu05}. It can be seen that our transformed $V$-band
magnitudes are consistent with those reported by \cite{chu05} to
within 0.1 mag, suggesting that our transformation to the
broad-band $V$ is reliable. This is consistent with expectations,
since there are no strong emission or absorption lines at the
$V$-band wavelength range for a typical SN II-P as demonstrated by
\citet{leo02}. Compared to SN 1999em, SN 2004dj shows a similar
evolution at the plateau phase, but a somewhat different evolution
in the nebular phase, particularly in the $R$ and $I$ bands. While
SN 1999em declines linearly in the $R$ and $I$ bands during the
nebular phase, SN 2004dj is flat in the $R$ band, and even
brightens slightly in the $I$ band. We emphasize, however, our
transformed $R$ and $I$ magnitudes for SN 2004dj may bear large
uncertainties due to the strong emission lines of H$\alpha$ and
the Ca~II IR triplet in our intermediate-band $i$ and $n$,
respectively. The different $R$ and $I$ evolution during the
nebular phase for SN 1999em and SN 2004dj thus does not
necessarily indicate a true difference between these two SNe.

\subsection{The Spectral Energy Distribution (SED) Evolution}

The spectral energy distribution (SED) of SN 2004dj can be best
studied by spectroscopy such as these done by \citet{kor05}.
Alternatively, a rough SED can be constructed from the observed
fluxes in various passbands at the same epoch. Because we have 12
intermediate-width passbands that cover 4000~\AA\, to 10000~\AA,
we can study the SED evolution of SN 2004dj from our observations.

Because of the number of filters involved to observe SN 2004dj, it
is not always possible to observe all bands in a single night, and
our definition of the same epoch refers to a reference date $\pm$
1 day. This is reasonable, considering the relatively slow
evolution of SN 2004dj (except during the transition from the
plateau to the nebular phase).

Figure 6 shows the SEDs of SN 2004dj at 61, 79, 112, 125, and 146
days after explosion. The SEDs are mostly a smooth function of
the wavelength, except in the $i$ and $n$ bands, which are
centered at 6700~\AA\, and 8500~\AA\, respectively. The deviation
of these two bands from the smooth distribution of the other
bands is likely to be caused by the strong lines in the spectrum
of SN 2004dj. Because of the small heliocentric recession
velocity (131 km s$^{-1}$) of NGC 2403, our $i$ filter is centered
at the H$\alpha$ emission, while the $n$ filter covers the strong
Ca~II triplet $\lambda\lambda$8498, 8542, 8662 line. From the
well-studied SN II-P 1999em, \citet{leo02} have shown that during
the late part of the plateau phase, a SN II-P is dominated by
strong lines with P-Cygni profiles. The emission component of
H$\alpha$ is generally much stronger than the absorption, as a
result, a net emission from H$\alpha$ when compared to the
neighboring continuum is expected. In Figure 6, this is verified
by the brighter magnitudes in the $i$ band than the $h$ and $j$
bands at 61 and 79 days after explosion. The absorption trough of
the Ca~II IR triplets, on the other hand, dominates the emission
component at the same phase. This is verified by the relatively
fainter magnitudes in the $n$ band when compared to the $m$ and
$o$ bands.

\citet{leo02} also show that during the nebular phase (later than
120 days after explosion), the spectrum of SN 1999em is dominated
by broad emissions. This is also true for the Ca~II IR triplet,
which shows multiple strong emission components but with little
absorption. Our SEDs for SN 2004dj closely follow this trend: the
$n$ band flux changes from a relative absorption at 61 and 79
days after explosion, to a relative emission at 125 and 146 days
after explosion. The smooth SED around $n$ band at 112 days after
explosion suggests that the absorption and emission components of
the Ca~II near-IR triplet is roughly balanced.

\subsection{Color evolution}

The $(V - R)$ (open circles) and $(V - I)$ (solid circles) color
curves of SN 2004dj are shown in Figure 7, together with
comparisons to those of SN 1999em \citep{leo02, ham01}. Before
$\sim$ 120 day after explosion, the two SNe show similar color
evolution [SN 2004dj has a slight bluer $(V - I)$ color]. In the
nebular phase, however, there is a relatively big difference
between the colors of the two SNe. SN 1999em remains nearly the
same color, while SN 2004dj becomes progressively redder. Several
factors may have contributed to this difference: there may be
large uncertainties in our transformed $R$- and $I$-band
magnitudes for SN 2004dj due to prominent emission lines of
H$\alpha$ and Ca~II near-IR triplet; the colors of SN 1999em may
have large uncertainties as well, since the photometry was not
performed after galaxy subtraction and background contamination
may be significant due to the faintness of SN 1999em in the
nebular phase.

Figure 7 also shows the evolution of SN 2004dj in two other
colors: $(d - g)$ (solid squares) and $(j - n)$ (solid stars). The
$d$ and $g$ bands, which are centered at $\sim$ 4540~\AA\, and
$\sim$ 5795~\AA, respectively, do not include strong lines of SNe
II-P. The $(d - g)$ color thus roughly equals the $(B - V)$ color
in the broad band. The $(B - V)$ color of SN 1999em and the $(d -
g)$ color of SN 2004dj both evolve from blue to red after
explosion, reach their reddest colors at $\sim$ 130-140 days, then
become progressively bluer [at a rate of $\sim$ 0.50 mag (100
days)$^{-1}$] as the SNe enter the nebular phase. The $(d - g)$
color of SN 2004dj is bluer than the $(B - V)$ color of SN 1999em
by $\sim$ 0.60$\pm$0.10 mag. This discrepancy may be caused by the
systematic difference between the intermediate- and broad-band
photometry. We were unable to convert our $d$-band photometry to
$B$ due to the lack of $c$-band (centered at $\sim$4220~\AA)
observations (cf. Eq. (2)).

The $(j - n)$ color of SN 2004dj shows a rapid and progressively
redder evolution. The $j$ band is centered at $\sim$ 7000 \AA\, and
roughly equals the broad $R$ band. It does not include strong
lines of a SN II-P. The $n$ band is centered at $\sim$ 8500 \AA\,
and roughly equals the broad $I$ band. This band, however,
includes the strong Ca~II IR triplet lines. Compared to the  $(R -
I)$ evolution of SN 1999em, the $(j - n)$ color of SN 2004dj
evolves toward red at a much faster  pace [$\sim$ 1.60 mag (100
days)$^{-1}$]. This is most likely caused by the brightening of
the strong Ca~II IR triplet emission in the $n$ band of SN 2004dj.

\section{Estimates of the explosion and the reddening}

\subsection{The explosion time}

The spectra taken immediately after the discovery indicated that
SN 2004dj was found long after the outburst (e.g., \citet{pat04},
\citet{kor05}). This is also supported by the photometric
observations. Our light curves of SN 2004dj (see Figure 4),
obtained $\sim$11 day after discovery, show that it was discovered
in the middle of the plateau phase. Following the method of
\citet{hen05}, we attempt to estimate the explosion time of SN
2004dj by comparing its light curve to those of other
well-observed SNe II-P. This method adjusts the time and magnitude
to find the best match using a $\chi^{2}$-minimizing technique.
Assuming SN 2004dj evolves like SN 1999em, the comparison of the
$V$-band data of the two SNe suggests an explosion date of
JD 2453168$\pm$1 (June 11$\pm$1, 2004) for SN 2004dj, with a
reduced $\chi^{2}$ of 1.39. This estimate is consistent with that
derived by \citet{chu05} (JD 2453170) from a comparison between
SN 2004dj and SN 1999gi. The uncertainty of our explosion date
($\pm$ 1 day) comes from the statistical error of the fitting only.

The explosion date derived above assume SN 2004dj has the same
duration of the plateau phase as SN 1999em. It was found from a
sample of 13 SNe II-P, however, the average plateau duration
varies by $\pm$21 days \citep{ham03, hen05}. As the error of the
plateau phase duration is propagated directly to the error of the
explosion date, we increase the uncertainty of our explosion date
to $\pm$ 21 days. The final explosion date, JD 2453168$\pm$21,
suggests that SN 2004dj was discovered at $\sim$50$\pm$21 days
after explosion.

\subsection{The interstellar extinction}

The Galactic extinction towards NGC 2403/SN 2004dj is known to be
$A^{\rm{gal}}_{V}=0.13$ mag \citep{sch98}, corresponding to a
color excess of $E(B -  V)$ = 0.04 mag (adopting the standard
reddening laws of \citet{car89}). However, it is rather difficult
to accurately measure the extinction towards SN 2004dj within its
host galaxy (NGC 2403).

With the assumption that all SNe II-P have the same intrinsic
color at the end of the plateau phase \citep{east96, ham04}, the
color evolution of SN 2004dj may provide clues to its reddening.
Since the early spectra of SN 2004dj are very similar to those of
SN 1999em \citep{pat04}, it is reasonable to use SN 1999em as a
comparison SN. We adopt a reddening of $E(B-V) = $ 0.10 mag for SN
1999em following the analysis by \citet{bar00} and \citet{ham01}.
A $\chi^{2}$-fitting algorithm was then used to compare the colors
of SN 2004dj with those of SN 1999em from 60 days to 130 days
after explosion. The $(V - R)$ fit yields a reddening of $E(B -
V) \sim 0$ mag for SN 2004dj, which is less than the Galactic
reddening of $E(B - V)$  = 0.04 mag. The $(V - I)$ fit yields a
negative reddening of $E(B -  V)$ = $-$0.19$\pm$0.05 mag for SN
2004dj (cf. Figure 6), which is physically unrealistic. Our
failure to derive a reasonable reddening for SN 2004dj using the
intrinsic color method suggests that the uncertainty of the method
may be large, as also noted by \citet{ham04}. We also note that
since our transformed $R$ and $I$ magnitudes for SN 2004dj are not
reliable, the reddening estimates from the $(V - R)$ and $(V - I)$
color evolution are questionable as well.

An alternative way to determine the reddening towards SN 2004dj is
to use the equivalent width (EW) of the Na I~D interstellar
absorption lines. The presence of narrow Na I~D absorption lines in
the early spectrum of SN 2004dj, with EW $\sim$ 1.1~\AA, was
reported by \citet{pat04}. Based on this EW, \citet{pat04}
estimated the reddening towards SN 2004dj within NGC 2403 is $E(B
- V)$ = 0.18 mag. The correlation between the EW of the Na I~D
interstellar absorption and the reddening, unfortunately, is
quite uncertain. Using EW = 1.1~\AA\, and the prescription by
\citet{bar90}, we derived a reddening of $E(B-V)$  = 0.28 mag for
SN 2004dj within NGC 2403. The prescription by \citet{mun97}, on
the other hand, suggests a host galaxy reddening of $E(B - V)$ =
0.40 mag. The average of the three measurements is $E(B - V)$ =
0.29$\pm$0.11 mag. When combined with the Galactic component, the
total reddening towards SN 2004dj is $E(B - V)$ = 0.33$\pm$0.11
mag, which is consistent with the estimate $E(B - V)$ =
0.35$\pm$0.05 mag from the spectral fit to the SED of S96
\citep{wxf05}.

We adopt $E(B-V)$ = 0.33$\pm$0.11 mag ($A_{V}$ = 1.02$\pm$0.34
mag) as our final reddening estimate for SN 2004dj. When this
correction for reddening is applied, we obtain the intrinsic peak
brightness for SN 2004dj as $M^{V}_{max} = - 16.70\pm0.35$ mag
(the major error is contributed from the uncertainty of the
reddening correction), which is typical for a SN II-P.

\section{The bolometric light curve and the mass of $^{56}$Ni}
\subsection{The bolometric light curve}
SNe II-P produce a large range of $^{56}$Ni masses from $\sim
10^{-3}$ to $> 10^{-1} M_{\odot}$. \citet{ham03} showed that a
correlation exists between the amount of $^{56}$Ni produced in the
explosion and the absolute magnitude in the $V$-band during the
plateau phase. Our observations suggest that SN 2004dj has a
plateau luminosity of $M^{V}_{p} = -16.55\pm0.35$ mag. We
therefore estimate that amount of $^{56}$Ni produced during the
SN 2004dj explosion to be intermediate between 0.01$M_{\odot}$ and
0.1$M_{\odot}$ if it follows the correlation shown in Figure 3 of
\citet{ham03}. To better estimate the mass of $ ^{56}$Ni produced
during the explosion, we construct the ``bolometric" light curve
for SN 2004dj by integrating the flux in the 12 intermediate
bands. The decline rate of the ``bolometric" light curve for SN
2004dj is smaller than that of SN 1987A [0.73$\pm$0.30 mag (100
days)$^{-1}$ vs. 1.06$\pm$0.06 mag (100 days)$^{-1}$]. This
pseudo-bolometric light curve may not be accurate due to our lack
of the ultraviolent and infrared photometry for SN 2004dj. The
resulting light curve, together with that of SN 1987A
\citep{sun90}, are shown in Figure 8.

\subsection{The ejected $^{56}$Ni mass}
\subsubsection{The $^{56}$Ni mass from the bolometric light curve}
The luminosity of a SN II-P during the nebular exponential decay
is controlled by the radioactive decay of the newly synthesized
materials during the SN explosion \citep{wea80}. If all the gamma
rays generated  by the $^{56}$Co $\rightarrow$ $^{56}$Fe decay are
fully thermalized, the decline rate of the light curve at the
nebular phase should equal the $^{56}$Co decay slope, and the mass
of $^{56}$Ni can be determined from the bolometric luminosity on
the exponential tail in Figure 8 using the equation below
\citep{ham03},
\begin{equation}
M_{\rm{Ni}} = 7.866\times10^{-44} L{}\exp[\frac{(t-t_{0})/(1+z)-
\tau_{\rm{Ni}}}{\tau_{\rm{Co}}}]M_{\odot},
\end{equation}
where $t_{0}$ is the explosion epoch, $\tau_{\rm{Ni} } = 6.1$ days
is the half-life of $^{56}$Ni and $\tau_{\rm{Co}} = 111.26$ days
is $e$-folding time of $^{56}$Co. $L$ is the tail luminosity
which is calculated by Eq.(1) of \citet{ham03}.

Using Eq.(3) we estimated $M_{\rm{Ni}}$ from each point on the
late-time tail for SN 2004dj. The average of these mass estimates
is $M_{\rm{Ni}}$ = 0.025$\pm$0. 010$M_{\odot}$, which is similar
to that produced in SN 1999em ($M_{\rm{Ni}}$ = 0.021$\pm$0.002
$M_{\odot}$; \citet{elm03b}). The error in the $^{56}$Ni mass is
caused by the uncertainties in the explosion date, extinction, and
the distance to NGC 2403.

\subsubsection{The $^{56}$Ni Mass from the ``Steepness  of Decline" Correlation}

A correlation between the maximum gradient at the transition phase
in the $V$ band and the photometric estimate of $M(^{56}\rm{Ni})$
has been reported by \citet{elm03a}. A characteristic steepness
parameter has been defined as \emph{S} = $-dM_{V}$/d$t$, which
corresponds to the maximum gradient rate of the light curve
between plateau and nebular phase. Based on a sample of ten SNe
II-P, they derived a linear relation between $M_{\rm{Ni}}$ (in
unit of solar mass $M_{\odot}$) and \emph{S},

\begin{equation}
\log\,M(^{56}\rm{Ni}) = -6.2295\,\emph{S} - 0.8147.
\end{equation}

Using the template flux from Eq.(1) of \citet{elm03a}, we fit the
light curves of SN 2004dj in several intermediate bands ($f$, $h$,
$j$, $k$ and $m$) that do not include strong emission or
absorption lines, as well as in the transformed $V$ band. The
results are shown in Figure 9. For each band, the upper panel
shows the best fit for the light curve, while the lower panel
shows the profiles of \emph{S} and the derived inflection time $t_{\rm{u}}$.
The average value of the ``steepness" parameter \emph{S} from fitting
the light curves is $0.141\pm0.005$, which suggests a $^{56}$Ni
mass of $0.020\pm0.002$ $M_{\odot}$. This value is consistent with
that determined from the bolometric light curve study, providing
support for the validity of the steepness parameter method.
However, one should note that in the case of SN 2003gd, \citet{hen05}
find the mass of $^{56}$Ni estimated from this method is significantly
lower than that from tail luminosity method.

The above two determinations allow us to estimate the amount of
$^{56}$Ni mass in SN 2004dj is 0.023$\pm0.005$ $M_{\odot}$. This
value is typical for a normal SN II-P, and is consistent with the
independent estimates from the H$\alpha$ luminosity or comparison
with SN 1987A \citep{chu05}.

\section{The progenitor star of SN 2004dj}

The detailed photometric observations may provide valuable
constraints on the explosion parameters of the supernova and hence
on the progenitor star \citep{hen05}. The hydrodynamical models of
SNe II-P light curves originally proposed by \citep{lina85}
predicted a correlation between the phenomenological parameters
(the plateau duration, the absolute magnitudes and photosphere
velocity at the middle of the plateau) and the physical parameters
such as the explosion energy $E_{exp}$, the ejecta mass $M_{ej}$,
and the initial radius of the pre-supernova star $R_{pSN}$.

Although the early observations immediately after the explosion
were unavailable for SN 2004dj, the fact that it resembles SN
1999em allows us to make an assumption about the length of its
plateau phase ($\Delta$t $\sim 80\pm21$ days). Adopting the
reddening estimate in $\S$ 4.2 and standard reddening law of
R$_{V}$ = 3.1, we derived the absolute magnitude during the
plateau phase as $M_{V} = -16.55\pm0.35$ mag. From the spectrum
observed at 52 days after explosion \citep{pat04}, we estimate a
photosphere velocity ($v_{\rm{ph}} \approx 3933\pm189$ km
s$^{-1}$) from weak Fe II absorption lines near 5000 \AA. With
these observed parameters, we can estimate the explosion
parameters by using the simple approximation formulae from
\citep{lina85}, which are here give as

\begin{equation}
\log\,(\frac{E_{\rm{exp}}}{10^{51}erg}) = 0.135\,M_{V} +
2.34\,\log\,(\frac{\Delta t}{days}) +
3.13\,\log\, (\frac{v_{\rm{ph}}}{10^3 km s^{-1}}) -4.205,\\
\end{equation}
\begin{equation}
\log\,(\frac{M_{\rm{ej}}}{M_{\odot}}) = 0.234\,M_{V} +
2.91\,\log\,(\frac{\Delta t}{days}) +
1.96\,\log\, (\frac{v_{\rm{ph}}}{10^{3}km s^{-1}}) -1.829,\\
\end{equation}
\begin{equation}
\log\,(\frac{R_{\rm{pSN}}}{R_{\odot}}) = -0.572\,M_{V} -
1.07\,\log\,(\frac{\Delta t}{days}) -
2.74\,\log\, (\frac{v_{\rm{ph}}}{10^{3}km s^{-1}}) -3.350.\\
\end{equation}

Combing the observed parameters and the above relations yields an
explosion energy $E_{exp}\approx
0.75^{+0.56}_{-0.38}\times10^{51}$ erg, an ejecta mass $M _{ej}
\approx 10.0^{+7.4}_{-5.2}$ $M_{\odot}$, and an initial radius
$R_{pSN} \approx 282^{+253}_{-122}$ $R_{\odot}$ for the
pre-supernova star of SN 2004dj. Taking into account approximately
2--3 $M_{\odot}$ for the remnant neutron star and the mass loss
lost by stellar wind prior to the SN explosion, we obtain a
possible range of 8--20 $M_{\odot}$ for the main sequence mass of
the progenitor, which is consistent with the estimates determined
by the turnoff mass of the star cluster hosting SN 2004dj
\citep{wxf05, mai04}. The derived parameters for the progenitor of
SN 2004dj suggest it was a K0--M0 supergiant \citep{cox00}.

\section{Conclusions}

This paper presents the photometry of SN 2004dj taken with 12
intermediate-band filters, and obtained from $\sim$ 60 to 200 days
after the explosion. Our observations show that SN 2004dj was
discovered in the middle of the plateau phase. The multicolor
light curves show a plateau phase of about 45 days, a transition
period of about 30 days during which the SN declines dramatically,
and a nebular phase during which the SN declines at a rate of
$\sim$ 0.10$\pm$0.03 mag (10 days)$^{-1}$ (consists with the decay
rate of $^{56}$Co).

The SEDs for SN 2004dj are constructed from the measured flux in
the 12 passbands. These SEDs show an evolution that is similar to
the spectral evolution of a normal SN II-P. A flux peak near 6600~\AA\,
is observed, which is consistent with the strong
H$\alpha$ emission seen in these objects. The flux around
8500~\AA\ evolves from a flux deficit to a flux peak, consisting
with the evolution of the strong Ca~II near-IR triplet line.

By comparing the light curve of SN 2004dj to the well observed SN
II-P 1999em, we estimate the explosion date for SN 2004dj to be JD
2453168$\pm$21 (June 11, 2004), approximately 50 day before the
actual discovery of SN 2004dj.

Finally, we estimate the $^{56}$Ni mass synthesized during the SN
2004dj explosion using two methods: the bolometric tail luminosity
and the ``steepness" parameter. These methods yield a value of
$M_{\rm{Ni}}$ = 0.023$\pm0.005$ $M_\odot$ for SN 2004dj, which is
typical for a normal SN II-P. Comparing our observed parameters of
SN 2004dj with the analytical models of SN II-P light curves
\citep{lina85}, We derive for SN 2004dj an explosion energy of
$0.75^{+0.56}_{-0.38}\times10^{51}$ erg, an ejecta mass of
$10.0^{+7.4}_{-5.2} M_{\odot}$, and an initial radius of
$282^{+253}_{-122} R_{\odot}$. These parameters suggest that the
progenitor of SN 2004dj is likely to be a K0--M0 type supergiant.

\acknowledgments This work was supported by National Science
Foundation of China (NSFC grant 10303002, 10473012 and 10573012),
National Key Basic Research Science Foundation (NKBRSF
TG199075402), and basic Research Foundation at Tsinghua University
(2005). The work of W.L. at U.C. Berkeley is supported by National
Science Foundation grant AST-0307894. We thank Dr. Yang Yanbin for
his helpful discussions on the reduction of the photometric data.

\clearpage
\begin{figure}
\includegraphics[angle=-90,width=160mm]{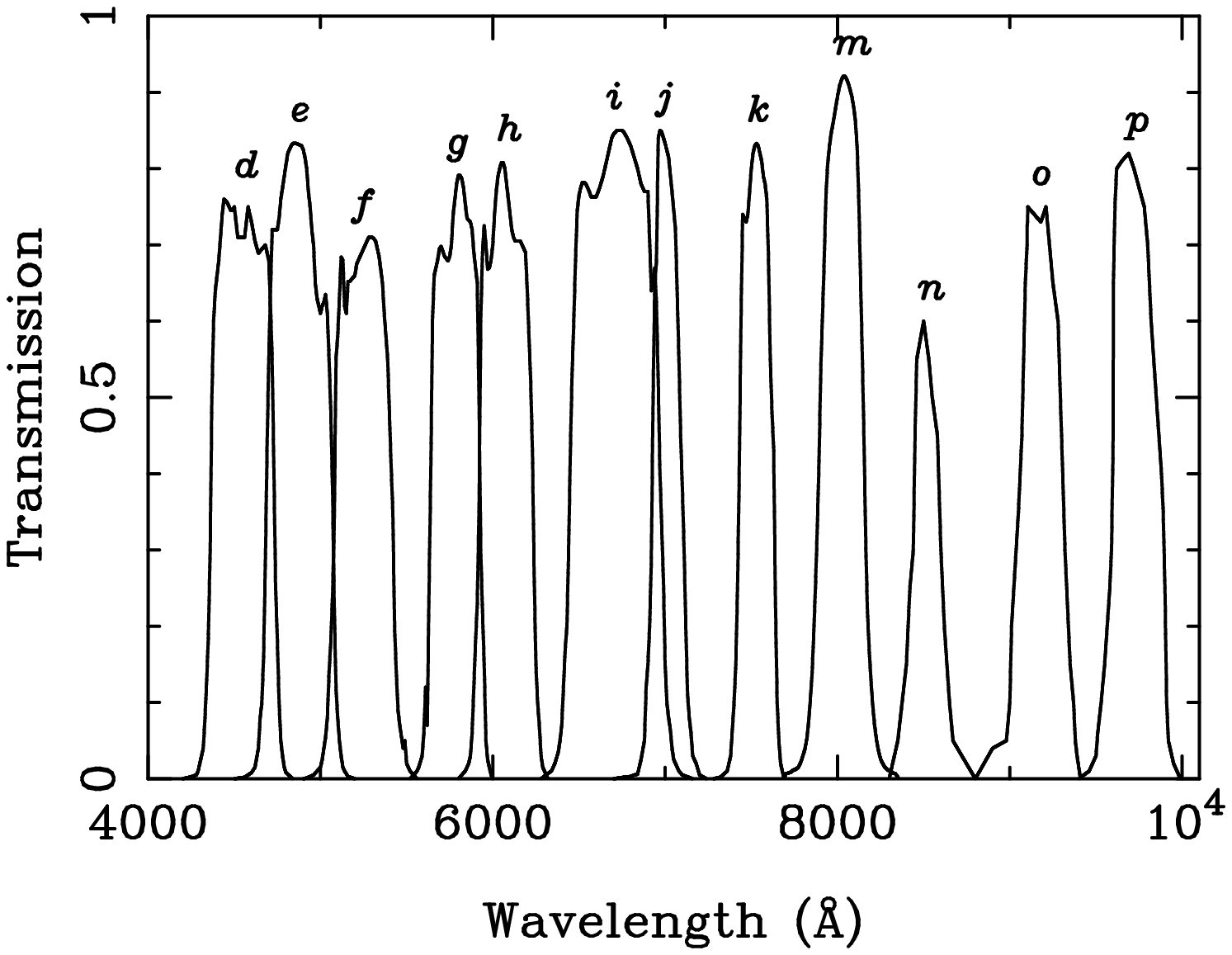}\caption{The transmission
curves of the 12 bands ($d-p$) in the BATC system. }
\label{fig:one}
\end{figure}

\clearpage
\begin{figure}
\includegraphics[angle=-90,width=160mm]{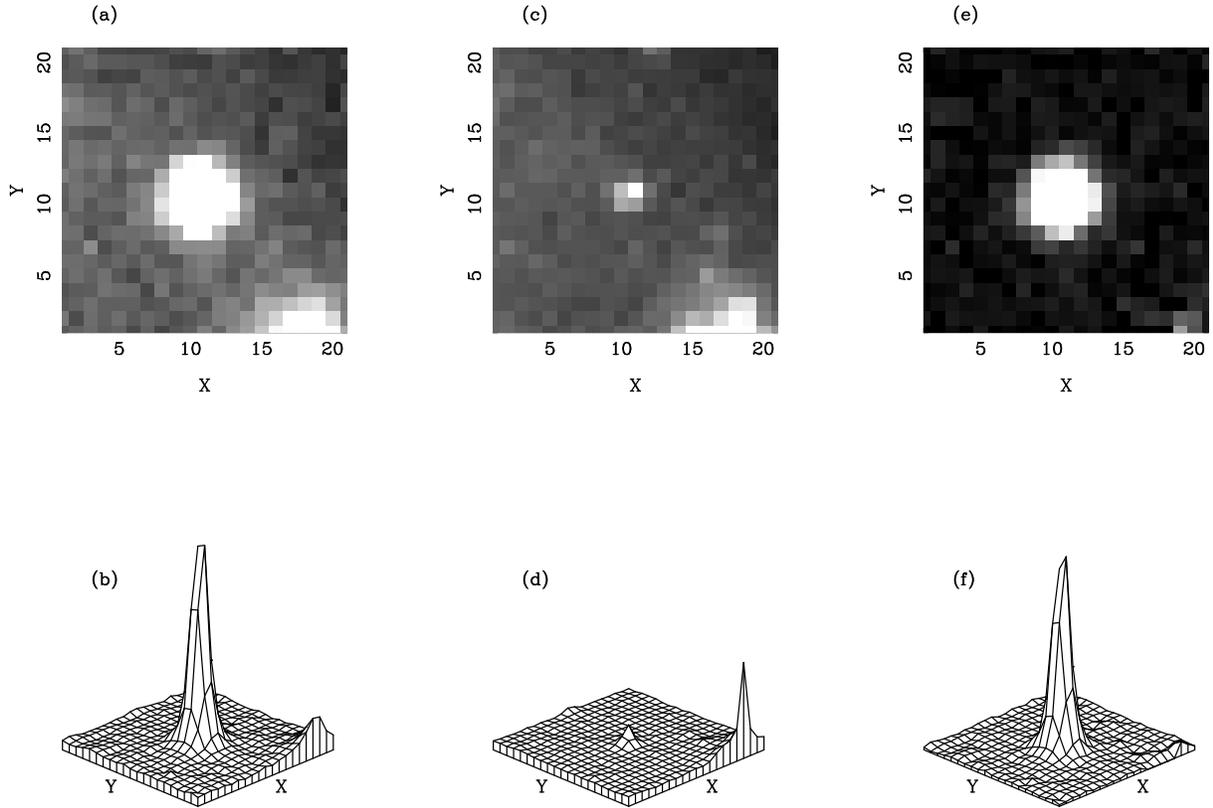}\caption{Demonstration of the
image subtraction process. $(a)$ The original image of SN 2004dj;
$(b)$ The surface plot for the original image; $(c)$ The template
image before SN 2004dj; $(d)$ The surface plot for the template
image; $(e)$ The image of SN 2004dj after galaxy subtracting,
notice the nearby H~II region is cleanly subtracted; $(f)$ the
surface plot for the subtracted image.} \label{fig:two}
\end{figure}

\clearpage
\begin{figure}
\includegraphics[angle=0,width=160mm]{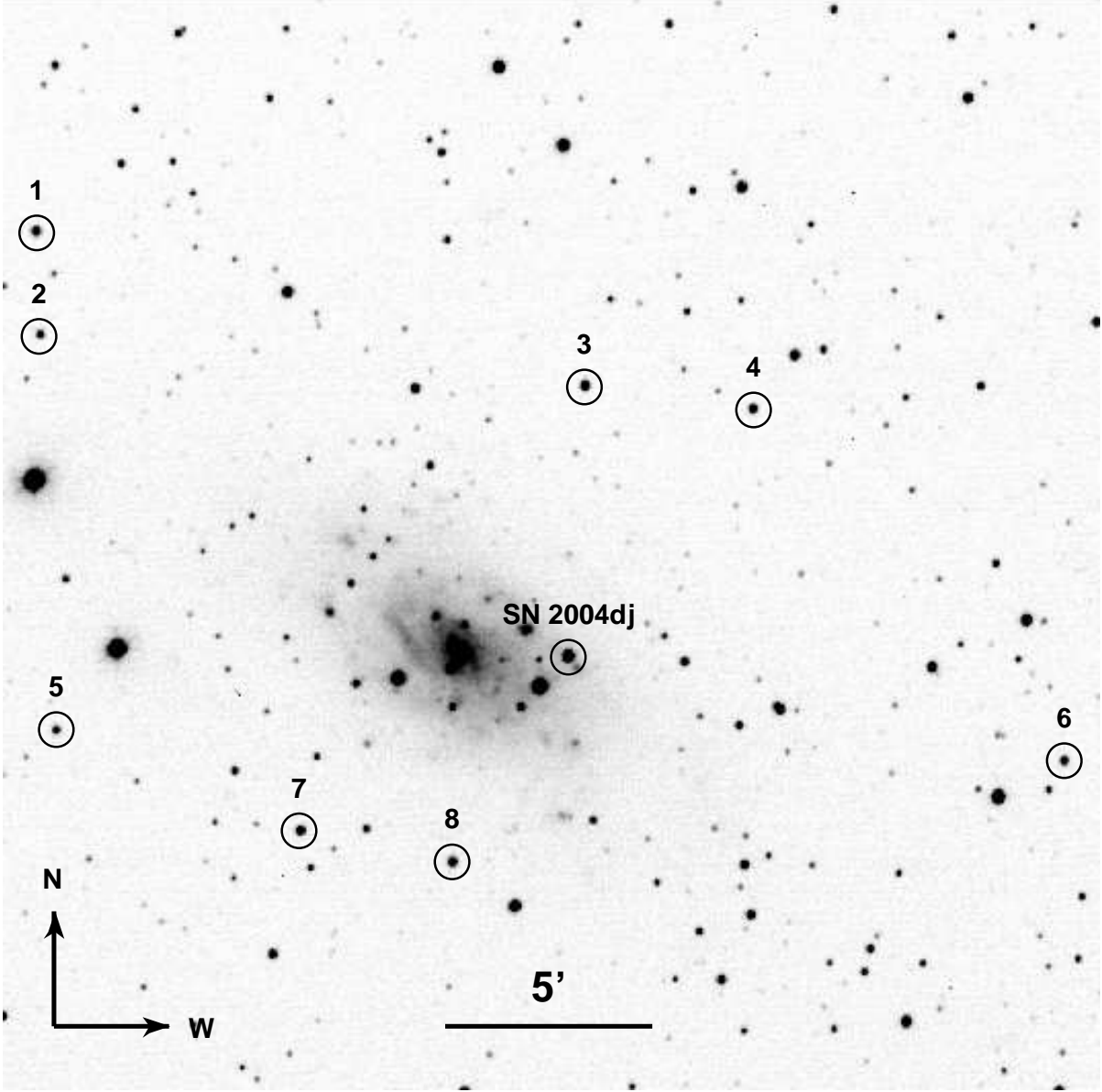}\caption{SN 2004dj in NGC 2403.
This is an $i$-band image taken with the BATC 60/90 cm Schmidt
telescope on 2004 August 12 UT, 13 days after discovery. The SN
and some reference stars are marked.} \label{fig:three}
\end{figure}

\clearpage
\begin{figure}
\includegraphics[angle=-90,width=160mm]{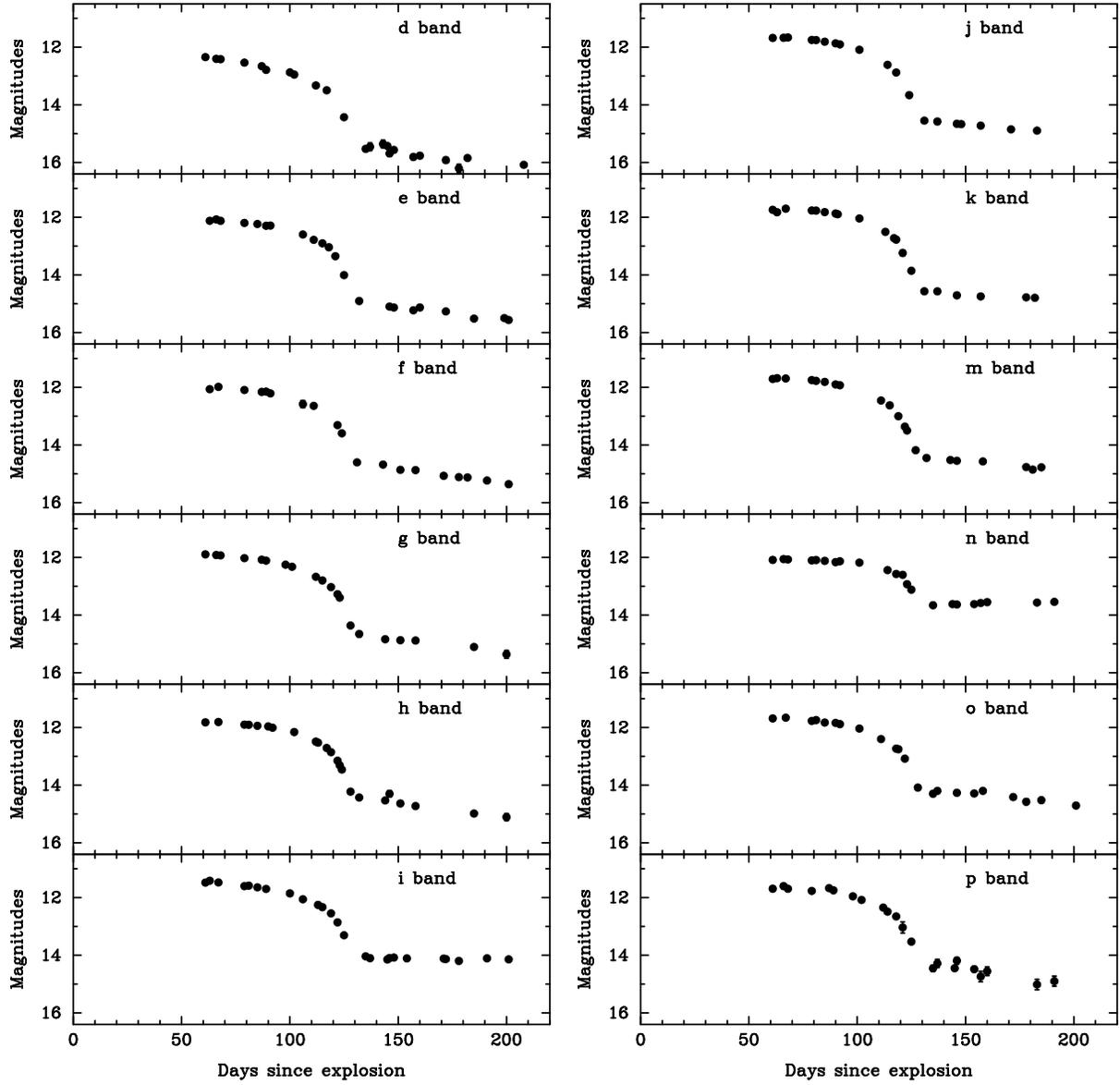}\caption{Light curves of SN 2004dj
in the 12 intermediate bands.} \label{fig:four}
\end{figure}

\clearpage
\begin{figure}
\includegraphics[angle=-90,width=160mm]{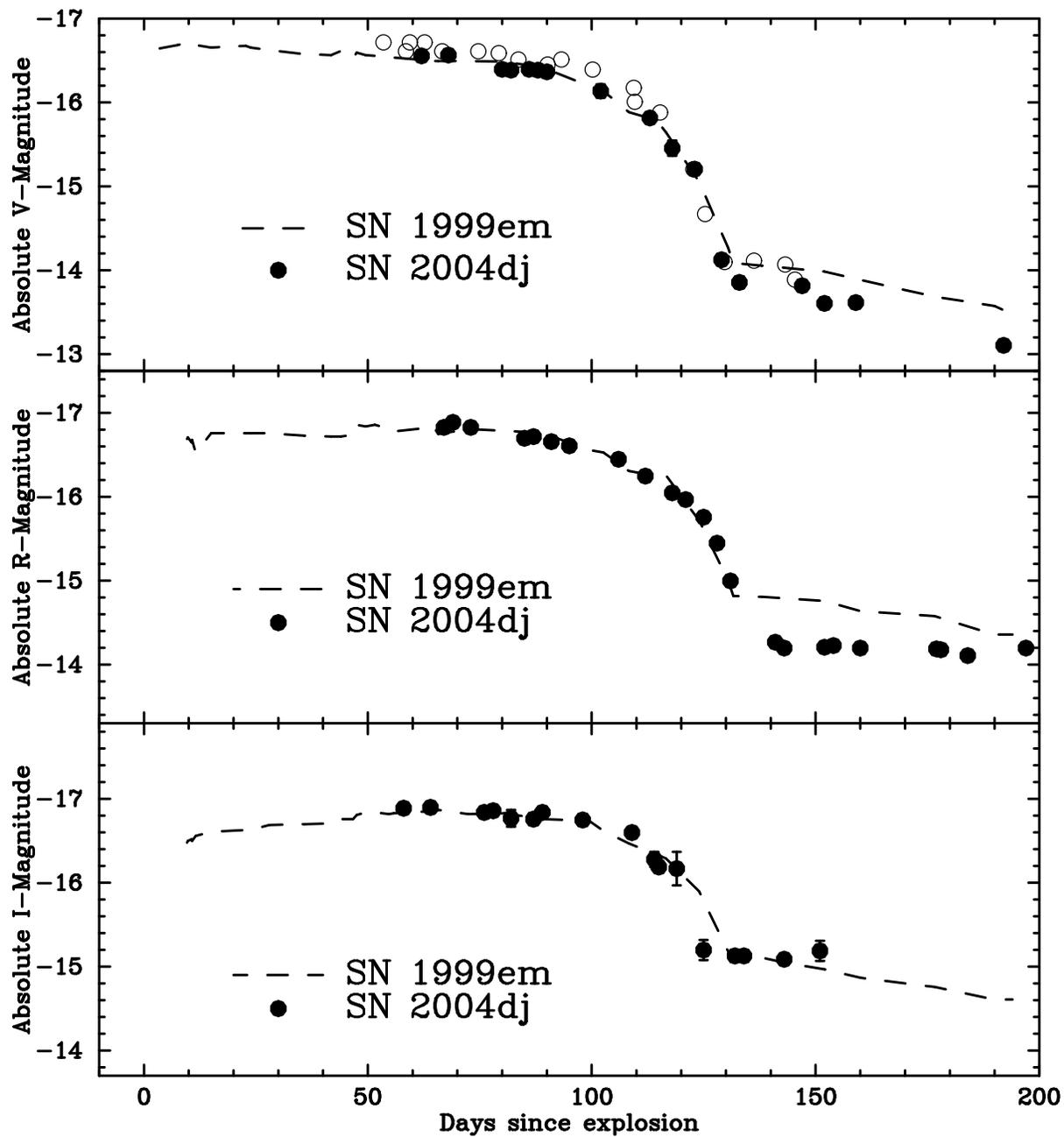}
\caption{Broad-band $VRI$ light curves of SN 2004dj (filled symbols). Also over-plotted are light curve of SN
1999em from \citet{ham01} and \citet{leo02}. The $V$-band magnitudes reported by \citet{chu05}
are shown as open symbols.} \label{fig:five}
\end{figure}

\clearpage
\begin{figure}
\includegraphics[angle=-90,width=160mm]{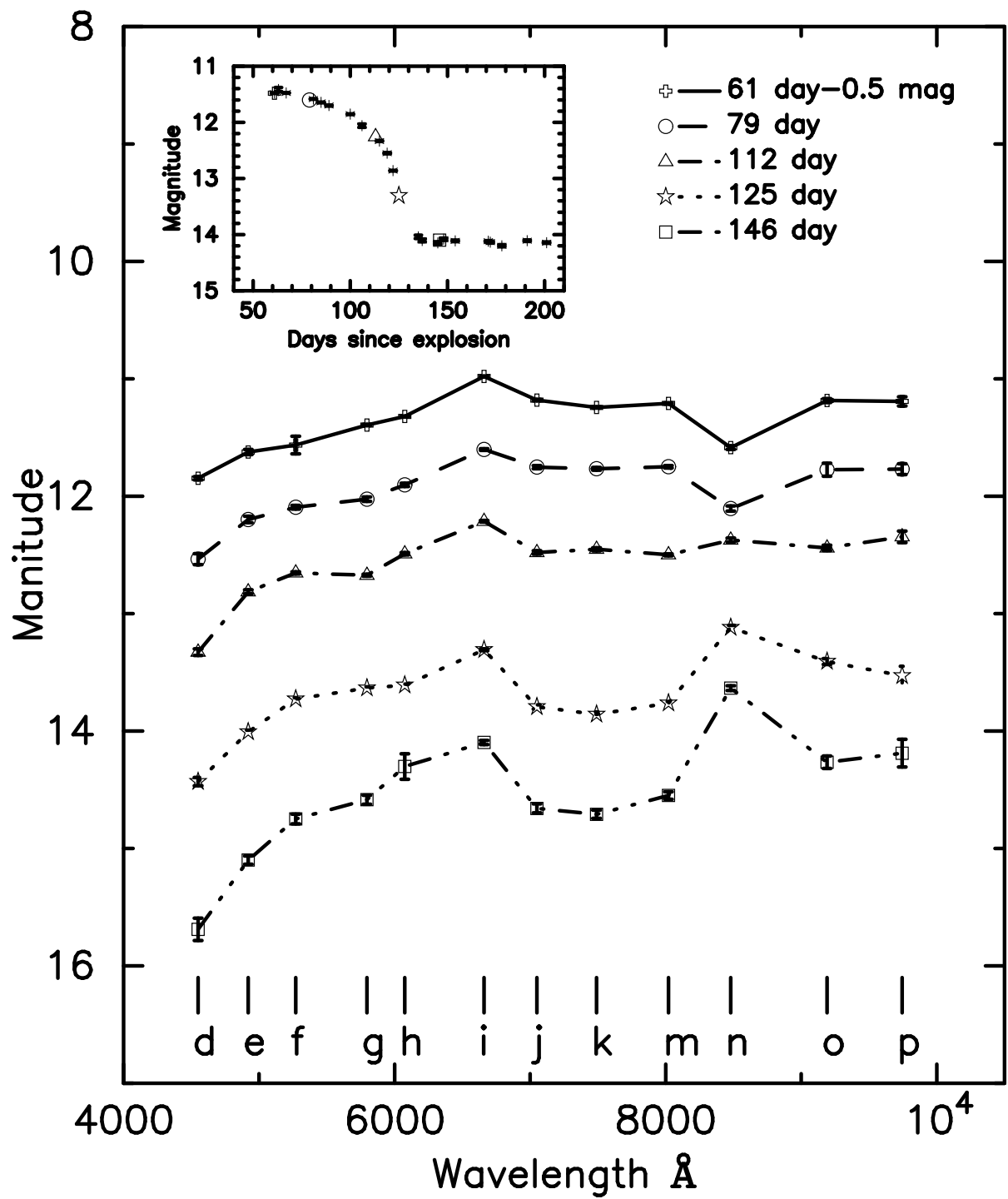}\caption{The spectral energy
distribution (SED) of SN 2004dj at days 61, 79, 112, 125, and 146 after explosion. The inset shows the location
of these SEDs in the $i$-band light curve.} \label{fig:six}
\end{figure}

\clearpage
\begin{figure}
\includegraphics[angle=0,scale=.6]{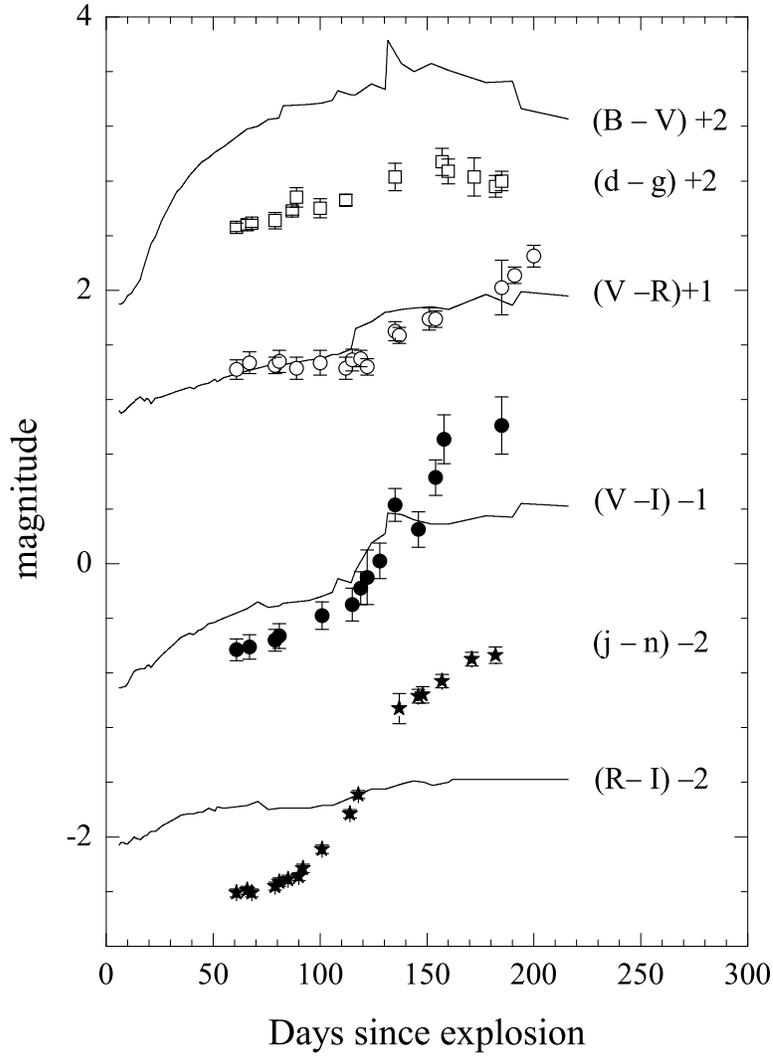}\caption{The color curves of SN
2004dj which have been arbitrarily shifted in magnitudes for clarity. The solid lines represent the data of SN
1999em from \citet{ham01} and \citet{leo02}.} \label{fig:seven}
\end{figure}

\clearpage
\begin{figure}
\includegraphics[angle=-90,width=160mm]{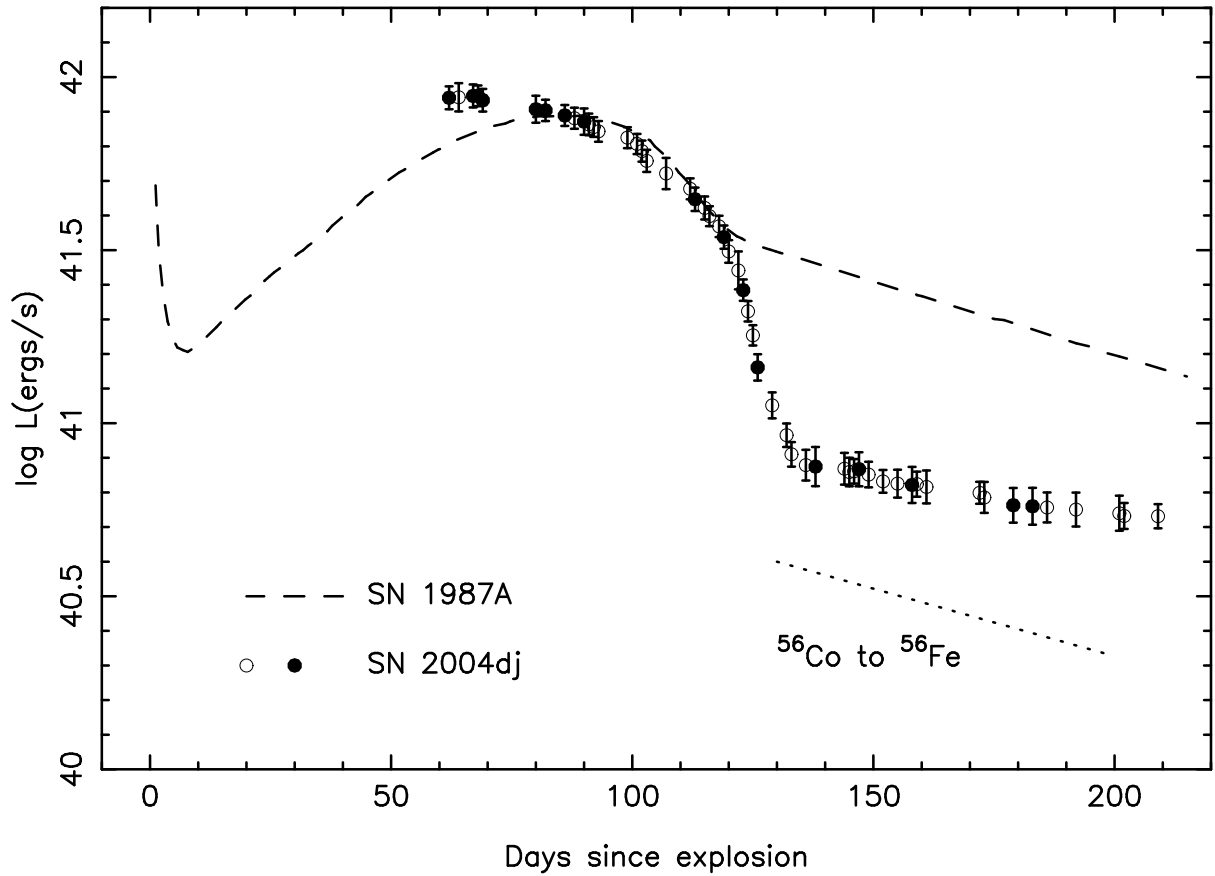}\caption{The bolometric light
curve of SN 2004dj compared to that of SN 1987A. The open circles are interpolated data points. The reported
errors are 1$\sigma$. The data of SN 1987A come from \citet{sun90}. The dotted line is the slope of $^{56}$Co to
$^{56}$Fe decay.} \label{fig:eight}
\end{figure}

\clearpage
\begin{figure}
\includegraphics[angle=-90,width=160mm]{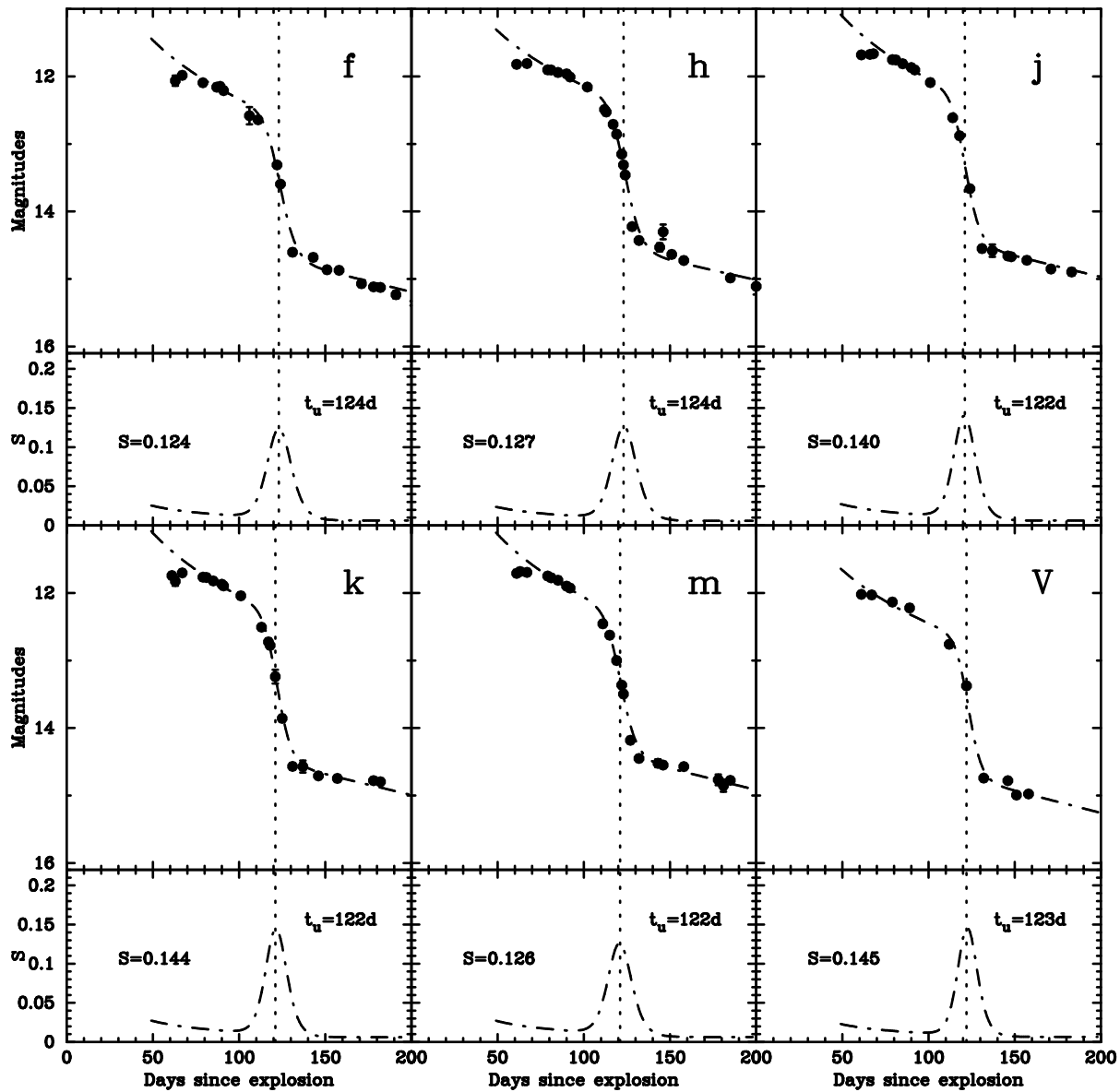}\caption{Determination of the
steepness parameter and the time of inflection for the $f$, $h$, $j$, $k$, $m$, and $V$ bands. For each
passband, the upper panel shows the light curve (solid dots) together with the best fit (dashed curve), while
the lower panel shows the profile for the steepness \emph{S} parameter. The inflection time $t_{\rm{u}}$ is marked by
a dotted line that corresponds to the maximum of \emph{S}.} \label{fig:night}
\end{figure}

\clearpage

\begin{deluxetable}{ccccccccccccccc}
\tabletypesize{\tiny}
\rotate
\tablecaption{Photometry of Comparison Stars\label{tbl-1}}
\tablewidth{0pt}
\tablehead{
\colhead{ID.} &
\colhead{$\alpha$(2000.0)} &
\colhead{$\delta$(2000.0)} &
\colhead{d} &
\colhead{e} &
\colhead{f} &
\colhead{g} &
\colhead{h} &
\colhead{i} &
\colhead{j} &
\colhead{k} &
\colhead{m} &
\colhead{n} &
\colhead{o} &
\colhead{p} \\}
\startdata
1&07:35:09.74&65:46:36.5&13.30(01)&13.12(01)&12.94(01)&12.85(01)&12.78(01)&12.68(01)&12.57(01)&12.58(01)&12.55(01)&12.56(01)&12.52(01)&12.52(02)\\
2&07:35:10.62&65:44:02.4&14.22(01)&14.06(01)&13.90(01)&13.79(01)&13.73(01)&13.58(01)&13.50(01)&13.51(01)&13.45(01)&13.46(01)&13.39(03)&13.42(03)\\
3&07:37:21.71&65:42:41.4&13.06(01)&12.94(01)&12.81(01)&12.77(01)&12.74(01)&12.79(01)&12.55(01)&12.62(01)&12.58(01)&12.57(01)&12.58(02)&12.62(02)\\
4&07:38:02.01&65:42:03.3&14.06(01)&13.77(01)&13.57(01)&13.35(01)&13.25(01)&13.06(01)&12.92(01)&12.89(01)&12.84(01)&12.81(01)&12.75(02)&12.76(02)\\
5&07:35:14.21&65:34:13.1&14.12(01)&13.88(01)&13.72(01)&13.57(01)&13.52(01)&13.36(01)&13.22(01)&13.27(01)&13.23(01)&13.18(01)&13.20(02)&13.22(03)\\
6&07:39:15.32&65:33:11.9&13.91(01)&13.71(01)&13.54(01)&13.44(01)&13.39(01)&13.26(01)&13.16(01)&13.18(01)&13.14(01)&13.13(01)&13.11(02)&13.14(03)\\
7&07:36:12.57&65:31:41.2&13.72(01)&13.42(01)&13.22(01)&13.04(01)&12.95(01)&12.84(01)&12.66(01)&12.67(01)&12.60(01)&12.58(01)&12.52(01)&12.51(02)\\
8&07:36:48.80&65:30:53.3&13.23(01)&13.05(01)&12.93(01)&12.83(01)&12.79(01)&12.67(01)&12.56(01)&12.58(01)&12.53(01)&12.53(01)&12.50(01)&12.50(02)\\
9&07:36:50.58&65:24:06.2&13.96(01)&13.82(01)&13.69(01)&13.63(01)&13.61(01)&13.48(01)&13.37(01)&13.42(01)&13.37(01)&13.33(01)&13.34(02)&13.38(03)\\
10&07:37:13.64&65:53:08.3&13.59(01)&13.41(01)&13.25(01)&13.11(01)&13.05(01)&12.93(01)&12.79(01)&12.83(01)&12.79(01)&12.78(01)&12.73(02)&12.79(02)\\
11&07:34:36.72&65:57:59.6&13.36(01)&13.18(01)&13.00(01)&12.89(01)&12.81(01)&12.71(01)&12.58(01)&12.61(01)&12.56(01)&12.59(01)&12.55(01)&12.58(02)\\
12&07:39:37.89&65:49:20.6&14.21(01)&14.02(01)&13.90(01)&13.82(01)&13.75(01)&13.64(01)&13.49(02)&13.55(01)&13.52(01)&13.50(02)&13.51(03)&13.43(03)\\
13&07:39:25.03&65:44:04.4&13.57(01)&13.39(01)&13.21(01)&13.09(01)&13.03(01)&12.92(01)&12.77(01)&12.82(01)&12.78(01)&12.80(01)&12.72(02)&12.85(02)\\
14&07:40:58.95&65:32:20.1&12.66(01)&12.56(01)&12.41(01)&12.33(01)&12.31(01)&12.28(01)&12.10(01)&12.16(01)&12.10(01)&12.11(01)&12.11(01)&12.12(01)\\
15&07:34:13.90&65:44:54.6&13.07(01)&12.93(01)&12.82(01)&12.76(01)&12.72(01)&12.64(01)&12.54(01)&12.59(01)&12.56(01)&12.58(01)&12.57(01)&12.58(02)\\
16&07:33:52.97&65:31:28.5&13.76(01)&13.49(01)&13.28(01)&13.11(01)&13.05(01)&12.88(01)&12.74(01)&12.74(01)&12.67(01)&12.65(01)&12.60(01)&12.55(03)\\
17&07:34:37.01&65:31:55.4&12.32(01)&12.18(01)&12.04(01)&11.94(01)&11.91(01)&12.07(01)&11.68(01)&11.73(01)&11.70(01)&11.69(01)&11.68(01)&11.69(01)\\
18&07:38:30.15&65:15:59.4&13.36(01)&13.19(01)&13.04(01)&12.90(01)&12.85(01)&12.71(01)&12.60(01)&12.63(01)&12.55(01)&12.56(01)&12.57(02)&12.54(02)\\
19&07:41:13.26&65:24:06.0&12.63(01)&12.56(01)&12.44(01)&12.40(01)&12.38(01)&12.33(01)&12.20(01)&12.25(01)&12.22(01)&12.24(01)&12.26(01)&12.24(01)\\
20&07:39:12.62&65:25:01.3&13.98(01)&13.73(01)&13.57(01)&13.33(01)&13.24(01)&13.06(01)&12.92(01)&12.94(01)&12.85(01)&12.84(01)&12.83(02)&12.79(02)\\
\enddata
\tablecomments{All quantities are magnitudes. Uncertainties
in the last two digits are indicated in parentheses. Uncertainties
which are less than 0.01 indicate as 0.01 in the table.}
\end{deluxetable}

\clearpage

\thispagestyle{empty}

\begin{deluxetable}{cclllllllllllllll}
\tabletypesize{\tiny}
\rotate
\tablecaption{SN 2004dj photometric observations\label{tbl-2}}
\tablewidth{0pt}
\tablehead{
\colhead{Date(UT)} &
\colhead{JD\tablenotemark{a}} &
\colhead{$d$} &
\colhead{$e$} &
\colhead{$f$} &
\colhead{$g$} &
\colhead{$h$} &
\colhead{$i$} &
\colhead{$j$} &
\colhead{$k$} &
\colhead{$m$} &
\colhead{$n$} &
\colhead{$o$} &
\colhead{$p$} &
\colhead{$V$} &
\colhead{$R$} &
\colhead{$I$} \\}
\startdata
12/08/04&3230&12.35(01)&\nodata&\nodata&11.89(01)&11.82(01)&11.48(01)&11.68(01)&11.74(01)&11.71(01)&12.09(02)&11.69(02)&11.69(04)&12.00(05)&11.58(06)&11.67(06)\\
14/08/04&3232&\nodata&12.12(02)&12.06(08)&\nodata&\nodata&11.42(04)&\nodata&11.83(07)&11.68(03)&\nodata&\nodata&\nodata&\nodata&11.52(04)&\nodata\\
17/08/04&3235&12.40(03)&12.08(01)&\nodata&11.92(01)&\nodata&\nodata&11.67(01)&\nodata&\nodata&12.06(02)&\nodata&11.60(04)&\nodata&\nodata&\nodata\\
18/08/04&3236&\nodata&\nodata&11.98(01)&\nodata&11.81(01)&11.48(01)&\nodata&11.70(01)&11.70(01)&\nodata&11.66(02)&\nodata&12.04(06)&11.58(06)&11.66(07)\\
19/08/04&3237&12.42(02)&12.13(01)&\nodata&11.93(01)&\nodata&\nodata&11.66(01)&\nodata&\nodata&12.07(02)&\nodata&11.69(04)&\nodata&\nodata&\nodata\\
30/08/04&3248&12.54(05)&12.20(03)&12.09(02)&12.03(02)&11.90(01)&11.60(01)&11.75(01)&11.77(01)&11.75(01)&12.11(02)&11.77(06)&11.77(05)&12.16(06)&11.71(05)&11.72(06)\\
01/09/04&3250&\nodata&\nodata&\nodata&\nodata&11.91(01)&11.58(01)&11.76(01)&11.77(01)&11.78(01)&12.09(02)&11.74(02)&\nodata&12.17(06)&11.69(06)&11.70(07)\\
05/09/04&3254&\nodata&12.24(01)&\nodata&\nodata&11.94(01)&11.65(01)&11.81(01)&11.82(01)&11.81(01)&12.12(01)&11.83(02)&\nodata&\nodata&11.75(06)&\nodata\\
07/09/04&3256&12.66(01)&\nodata&12.16(01)&12.08(01)&\nodata&\nodata&\nodata&\nodata&\nodata&\nodata&\nodata&11.67(03)&\nodata&\nodata&\nodata\\
09/09/04&3258&12.79(07)&12.30(03)&12.15(02)&12.11(02)&11.96(01)&11.70(01)&\nodata&\nodata&\nodata&\nodata&\nodata&11.75(05)&12.23(05)&11.80(06)&\nodata\\
10/09/04&3259&\nodata&\nodata&\nodata&\nodata&\nodata&\nodata&11.87(01)&11.87(01)&11.90(01)&12.16(02)&11.84(03)&\nodata&\nodata&\nodata&11.80(07)\\
11/09/04&3260&\nodata&12.29(01)&12.21(01)&\nodata&\nodata&\nodata&\nodata&11.90(01)&\nodata&\nodata&\nodata&\nodata&\nodata&\nodata&\nodata\\
12/09/04&3261&\nodata&\nodata&\nodata&\nodata&12.01(01)&\nodata&11.90(01)&\nodata&11.93(01)&12.13(01)&11.88(02)&\nodata&\nodata&\nodata&11.72(07)\\
18/09/04&3267&\nodata&\nodata&\nodata&12.25(01)&\nodata&\nodata&\nodata&\nodata&\nodata&\nodata&\nodata&11.96(03)&\nodata&\nodata&\nodata\\
20/09/04&3269&12.87(02)&\nodata&\nodata&\nodata&\nodata&11.85(01)&\nodata&\nodata&\nodata&\nodata&\nodata&\nodata&\nodata&11.96(06)&\nodata\\
21/09/04&3270&\nodata&\nodata&\nodata&12.33(01)&\nodata&\nodata&12.09(01)&12.04(02)&\nodata&12.18(01)&12.04(02)&\nodata&12.46(08)&\nodata&11.84(07)\\
22/09/04&3271&12.95(02)&\nodata&\nodata&\nodata&12.16(01)&\nodata&\nodata&\nodata&\nodata&\nodata&\nodata&12.08(04)&\nodata&\nodata&\nodata\\
26/09/04&3275&\nodata&12.60(08)&12.58(13)&\nodata&\nodata&12.06(03)&\nodata&\nodata&\nodata&\nodata&\nodata&\nodata&\nodata&12.16(06)&\nodata\\
01/10/04&3280&\nodata&12.78(02)&12.64(01)&\nodata&12.49(01)&\nodata&\nodata&\nodata&12.46(01)&\nodata&12.40(02)&\nodata&\nodata&\nodata&\nodata\\
02/10/04&3281&13.33(03)&\nodata&\nodata&12.67(01)&12.53(01)&12.25(01)&\nodata&12.51(01)&\nodata&\nodata&\nodata&12.35(05)&12.78(06)&12.35(06)&11.96(08)\\
04/10/04&3283&\nodata&\nodata&\nodata&\nodata&\nodata&\nodata&12.61(02)&\nodata&\nodata&12.44(02)&\nodata&12.49(05)&\nodata&\nodata&\nodata\\
05/10/04&3284&\nodata&12.90(01)&\nodata&12.80(01)&\nodata&12.33(01)&\nodata&\nodata&12.63(01)&\nodata&\nodata&\nodata&12.92(06)&12.43(06)&12.21(10)\\
07/10/04&3286&13.50(03)&\nodata&\nodata&\nodata&12.71(01)&\nodata&\nodata&12.73(02)&\nodata&\nodata&\nodata&\nodata&\nodata&\nodata&\nodata\\
08/10/04&3287&\nodata&13.04(02)&\nodata&\nodata&\nodata&\nodata&12.88(01)&12.78(01)&\nodata&12.57(01)&12.73(02)&12.65(05)&\nodata&\nodata&12.37(06)\\
09/10/04&3288&\nodata&\nodata&\nodata&13.03(03)&12.86(02)&12.55(01)&\nodata&\nodata&13.00(02)&\nodata&12.75(03)&\nodata&13.15(06)&12.65(05)&12.32(10)\\
11/10/04&3290&\nodata&13.35(02)&\nodata&\nodata&\nodata&\nodata&\nodata&13.24(10)&\nodata&12.61(07)&\nodata&13.04(20)&\nodata&\nodata&\nodata\\
12/10/04&3291&\nodata&\nodata&13.31(01)&13.28(01)&13.15(01)&12.86(01)&\nodata&\nodata&13.36(01)&\nodata&13.08(03)&\nodata&13.39(06)&12.96(05)&12.49(20)\\
13/10/04&3292&\nodata&\nodata&\nodata&13.39(01)&13.31(01)&\nodata&\nodata&\nodata&13.50(01)&12.93(02)&\nodata&\nodata&\nodata&\nodata&\nodata\\
14/10/04&3293&\nodata&\nodata&13.60(01)&\nodata&13.46(01)&\nodata&13.66(02)&\nodata&\nodata&\nodata&\nodata&\nodata&\nodata&\nodata&\nodata\\
15/10/04&3294&14.43(04)&14.01(02)&\nodata&\nodata&\nodata&13.31(01)&\nodata&13.86(02)&\nodata&13.12(02)&\nodata&13.53(08)&\nodata&13.41(05)&\nodata\\
18/10/04&3297&\nodata&\nodata&\nodata&14.36(03)&14.23(02)&\nodata&\nodata&\nodata&14.18(03)&\nodata&14.09(08)&\nodata&14.47(06)&\nodata&13.45(12)\\
21/10/04&3300&\nodata&\nodata&14.60(03)&\nodata&\nodata&\nodata&14.55(04)&14.57(04)&\nodata&\nodata&\nodata&\nodata&\nodata&\nodata&\nodata\\
22/10/04&3301&\nodata&14.91(04)&\nodata&14.66(04)&14.43(03)&\nodata&\nodata&\nodata&14.45(04)&\nodata&\nodata&\nodata&14.74(07)&\nodata&\nodata\\
25/10/04&3304&15.53(06)&\nodata&\nodata&\nodata&\nodata&14.04(03)&\nodata&\nodata&\nodata&13.66(02)&14.30(07)&14.45(11)&14.86(10)&14.14(05)&13.43(07)\\
27/10/04&3306&15.46(14)&\nodata&\nodata&\nodata&\nodata&14.10(02)&14.58(10)&14.57(09)&\nodata&\nodata&14.20(07)&14.29(14)&\nodata&14.21(02)&13.43(07)\\
02/11/04&3312&15.36(14)&\nodata&14.68(05)&\nodata&\nodata&\nodata&\nodata&\nodata&14.52(06)&\nodata&\nodata&\nodata&\nodata&\nodata&\nodata\\
03/11/04&3313&\nodata&\nodata&\nodata&14.84(07)&14.53(10)&\nodata&\nodata&\nodata&\nodata&13.62(02)&\nodata&\nodata&14.95(13)&\nodata&\nodata\\
04/11/04&3314&15.43(04)&\nodata&\nodata&\nodata&\nodata&14.15(02)&\nodata&\nodata&\nodata&\nodata&\nodata&14.45(06)&\nodata&\nodata&\nodata\\
05/11/04&3315&15.69(10)&15.10(04)&\nodata&14.59(04)&14.30(03)&14.10(02)&14.66(04)&14.71(04)&14.55(03)&13.63(02)&14.27(05)&14.19(12)&14.72(06)&14.20(05)&13.47(06)\\
07/11/04&3317&15.57(07)&15.13(03)&\nodata&\nodata&\nodata&14.08(02)&14.67(04)&\nodata&\nodata&\nodata&\nodata&\nodata&\nodata&14.18(05)&\nodata\\
10/11/04&3320&\nodata&\nodata&14.86(04)&14.87(03)&14.64(02)&\nodata&\nodata&\nodata&\nodata&\nodata&\nodata&\nodata&14.99(07)&\nodata&\nodata\\
13/11/04&3223&\nodata&\nodata&\nodata&\nodata&\nodata&14.11(01)&\nodata&\nodata&\nodata&13.62(01)&14.29(07)&14.48(09)&15.00(06)&14.21(01)&13.37(12)\\
16/11/04&3326&15.81(09)&15.23(04)&\nodata&\nodata&\nodata&\nodata&14.72(04)&14.75(04)&\nodata&13.58(02)&\nodata&14.74(18)&\nodata&\nodata&\nodata\\
17/11/04&3327&\nodata&\nodata&14.87(03)&14.88(04)&14.73(03)&\nodata&\nodata&\nodata&14.57(03)&\nodata&14.20(05)&\nodata&14.98(06)&\nodata&13.07(16)\\
19/11/04&3329&15.77(08)&15.13(04)&\nodata&\nodata&\nodata&\nodata&\nodata&\nodata&\nodata&13.55(02)&\nodata&14.56(15)&\nodata&\nodata&\nodata\\
30/11/04&3340&\nodata&\nodata&15.07(03)&\nodata&\nodata&14.12(01)&14.85(03)&\nodata&\nodata&\nodata&\nodata&\nodata&\nodata&14.22(05)&\nodata\\
01/12/04&3341&15.92(10)&15.27(07)&\nodata&\nodata&\nodata&14.13(02)&\nodata&\nodata&\nodata&\nodata&14.41(09)&\nodata&\nodata&14.23(06)&\nodata\\
07/12/04&3347&16.21(14)&\nodata&15.12(03)&\nodata&\nodata&14.20(02)&\nodata&14.78(04)&14.77(08)&\nodata&14.58(10)&\nodata&\nodata&14.30(05)&\nodata\\
11/12/04&3351&15.85(06)&\nodata&15.13(03)&\nodata&\nodata&\nodata&14.90(05)&14.80(04)&14.85(09)&13.57(02)&\nodata&15.02(18)&\nodata&\nodata&\nodata\\
14/12/04&3354&\nodata&15.52(07)&\nodata&15.11(05)&14.99(03)&\nodata&\nodata&\nodata&14.78(04)&\nodata&14.52(10)&\nodata&15.22(06)&\nodata&13.21(20)\\\tablebreak
20/12/04&3360&\nodata&\nodata&15.23(05)&\nodata&\nodata&14.11(01)&\nodata&\nodata&\nodata&13.54(02)&\nodata&14.90(18)&15.32(06)&14.21(06)&\nodata\\
29/12/04&3369&\nodata&15.50(08)&\nodata&15.36(14)&15.11(12)&\nodata&\nodata&\nodata&\nodata&\nodata&\nodata&\nodata&15.47(06)&\nodata&\nodata\\
30/12/04&3370&\nodata&15.57(05)&15.36(03)&\nodata&\nodata&14.14(01)&\nodata&\nodata&\nodata&\nodata&14.71(07)&\nodata&\nodata&14.25(06)&\nodata\\
06/01/05&3377&16.08(07)&\nodata&\nodata&\nodata&\nodata&\nodata&\nodata&\nodata&\nodata&\nodata&\nodata&\nodata&\nodata&\nodata&\nodata\\
\enddata
\tablenotetext{a}{2450000+}
\end{deluxetable}

\clearpage
\begin{deluxetable}{ccr}
\tablecaption{The decline rates of different bands on the
exponential phase.\label{tbl-3}} \tablewidth{0pt} \tablehead{
\colhead{Filter name}&$\lambda$(\AA)&\colhead{Decline rate}\\
\colhead{}&\colhead{}& [mag (10 days)$^{-1}$]} \startdata
d&4540&$0.10\pm0.01$\\
e&4925&$0.08\pm0.01$\\
f&5270&$0.11\pm0.01$\\
g&5795&$0.11\pm0.01$\\
h&6075&$0.11\pm0.01$\\
i&6656&$0.00\pm0.01$\\
j&7057&$0.07\pm0.01$\\
k&7546&$0.06\pm0.02$\\
m&8023&$0.07\pm0.01$\\
n&8480&$-0.02\pm0.01$\\
o&9182&$0.08\pm0.01$\\
p&9739&$0.13\pm0.03$\\
\enddata
\end{deluxetable}

\end{document}